\documentclass[12pt, a4paper] {article}

\usepackage{amsfonts}
\usepackage{verbatim}
\usepackage{color}
\usepackage{amsmath}
\usepackage{graphicx}
\usepackage{bm}
\usepackage{amssymb}
\usepackage{latexsym}
\usepackage{mathrsfs}
\usepackage{textcomp}
\begin{document}

\begin{center}
{\bf\large{Symplectic gauge invariant reformulation of a free particle system on toric geometry} }

\vskip 1.5 cm

{\sf{ \bf Anjali S and Saurabh Gupta}}\\
\vskip .1cm
{\it Department of Physics, National Institute of Technology Calicut,\\ Kozhikode - 673 601, Kerala, India}\\
\vskip .15cm
{E-mails: {\tt anjalisujatha28@gmail.com, saurabh@nitc.ac.in}}
\end{center}
\vskip 1cm

\noindent
{\bf Abstract:}
We deduce the constraint structure and subsequently quantize a free particle system residing on a torus satisfying the geometric constraint $(r-a)=0$, within the framework of modified Faddeev-Jackiw formalism. Further, we reformulate the system as a gauge theory by the means of symplectic gauge invariant formalism. Finally, we establish the off-shell nilpotent and absolutely anti-commuting (anti-)BRST symmetries of the reformulated gauge invariant theory.

\vskip 1.5 cm

\noindent    
{\bf PACS}: 11.10.Ef, 11.15.-q, 11.30.-j

\vskip 1 cm
\noindent
{\bf Keywords}: Modified Faddeev-Jackiw formalism; Symplectic gauge invariant formalism; Constrained system; (anti-)BRST symmetries.

\newpage

\section{Introduction}

Gauge theories play a vital role in describing the fundamental interactions of nature \cite{int1}. Although the gauge theories are conceptualized beautifully yet the passage between configuration space and phase space  is not in one-to-one manner due to the presence of constraints \cite{cons, int4}.  The Dirac formalism provides a consistent and systematic treatment of constraints by categorizing them  as primary, secondary, tertiary, etc. and further into first-class and second-class \cite{int4}. 
The tedious algebraic calculations involved in the study of constrained systems by Dirac formalism led to an alternative methodology known as Faddeev-Jackiw formalism \cite{2, FJ1} which rely on the symplectic nature of phase space. The Faddeev-Jackiw formalism is further  modified to include the consistency condition of constraints in a Dirac way \cite{modFJ2} and it finds applications in many systems such as four dimensional BF theory \cite{modFJ_BF}, anti self-dual Yang-Mills equations \cite{modFJ3} and Christ-Lee model \cite{modFJ4} are the few names.

The importance of gauge theory attracted many to reformulate the second-class constrained theory as a gauge invariant one in order to unravel the gauge symmetries hidden in the system. The Batalin-Fradkin-Fradkina-Tyutin  formalism  is one such technique that relies on enlarging the phase space with the introduction of Wess-Zumino variables \cite{bfft2, bfft3}. 
Another famous formalism is due to Mitra and Rajaraman which deals with the systems possessing purely second-class constraints  \cite{mit1}.  Furthermore, an another methodology known as symplectic gauge invariant formalism \cite{4, sym}, based on the symplectic framework, is proposed which deals with the introduction of an arbitrary function depending upon the phase space as well as Wess-Zumino variables.

As far as the quantization of such gauge theories is concerned Becchi-Rouet-Stora-Tyutin (BRST) formalism provides a most intuitive approach\cite{brst1, brst3}. In this formalism, the notion of gauge invariance is replaced by the BRST invariance via enlarging the Hilbert space and the effects of gauge fixing are achieved without breaking the BRST invariance of the theory \cite{rotor}.

 On the other hand toric geometry provides a suitable framework in realizing many Calabi-Yau manifolds and their mirrors in the regime of string theory \cite{toric_str}. It also finds application in the discussion of geometric properties of manifolds in F-theory dualities \cite{toric_Fduality}. Whereas in the domain of quantum field theories, a free particle system on toric geometry has been explored within the Batalin-Fradkin-Vilkovisky framework \cite{6}. 
Moreover, its off-shell nilpotent and absolutely anti-commuting (anti-)BRST symmetries have been established within the framework of superfield formalism \cite{1, sf2} and the same system has been shown to behave as a model for the double Hodge theory \cite{hodge}.

In the present endeavor, one of our prime motives is to explore a free particle system on toric geometry with a geometrically motivated approach to deduce the constraint structure and subsequently carry out its quantization. Second key motive is to reformulate the system as a gauge theory with the aid of symplectic gauge invariant formalism and give an account for the gauge symmetries. Finally, we intend to provide (anti-)BRST symmetries for the newly reformulated gauge theory.

The contents of the paper are organized as follows. Section 2 gives a brief preamble about a free particle restricted to move on a torus, satisfying the geometric constraint $(r-a)=0$. Section 3 deals with constraint analysis and quantization of the system in modified Faddeev-Jackiw formalism. Our section 4 contains the reformulation of the system as a gauge theory by employing symplectic gauge invariant formalism. We derive the off-shell nilpotent and absolutely anti-commuting (anti-)BRST symmetries and the corresponding conserved charges, in section 5. Finally, in section 6, we provide  concluding remarks.

\section{Preliminaries}
The parametric equations of a torus, with axial circle in the $x-y$ plane centered at the origin of radius $R$ and having circular cross section of radius $r$, are given as 
\begin{equation} \label{p_torus}
	x \;=\; (R+r \sin\theta)\cos\phi, \quad
	y \;=\; (R+r \sin\theta)\sin\phi, \quad
	z \;=\; r \cos\theta,
\end{equation}
where the coordinates $\theta$ and $\phi$ ranges from $0$ to $2\pi$. In toric geometry, the coordinates $(r, \theta, \phi)$ satisfy the relation $(\sqrt{x^{2} + y^{2}} - R)^{2} + z^{2} = r^{2}$. 
Let us consider a free particle of unit mass ($m = 1$) 
which is restricted to move on a torus and is subjected to satisfy the  
geometric constraint $(r-a) = 0$ \cite{1}. In toric geometry \eqref{p_torus}, the Lagrangian describing this constrained particle is given as \cite{1}
\begin{equation} \label{L_original}
	L \;=\; \frac{1}{2}\dot{r}^{2} + \frac{1}{2}r^{2}\dot{\theta}^{2} + \frac{1}{2}(R+r \sin\theta)^{2}\dot{\phi}^{2} + \lambda(r-a),
\end{equation}
where $\dot{r}$, $\dot{\theta}$ and $\dot{\phi}$ are generalized velocities and $\lambda$ is a Lagrange multiplier. The canonical momenta corresponding to the  variables $r, \theta, \phi, \lambda$ are, respectively, given as
\begin{equation}
	P_{r} \;=\; \dot{r}, \quad P_{\theta} \;=\; r^{2}\dot{\theta}, \quad P_{\phi} \;=\; (R+r \sin\theta)^{2}\dot{\phi},\quad P_{\lambda} \;=\; 0.
\end{equation} 
The canonical Hamiltonian obtained from the Lagrangian \eqref{L_original} has following form 
\begin{equation} \label{H_c}
	H_{c} \;=\; \frac{P_{r}^{2}}{2} + \frac{P_{\theta}^{2}}{2r^{2}} + \frac{P_{\phi}^{2}}{2(R+r \sin\theta)^{2}} - \lambda(r-a).
\end{equation}
The constraint structure of the system can be deduced, using Dirac formalism, as
\begin{equation} \label{df_cons}
\varphi_{1} \;=\; P_{\lambda} \;\approx \; 0, \quad \varphi_{2} \;=\; (r-a) \; \approx \; 0, \quad \varphi_{3} \;=\; P_{r} \; \approx \; 0.
\end{equation}
Among these three constraints ($\varphi_{k}; k=1,2,3$), the non-vanishing Poisson bracket is $\left\lbrace \varphi_{2}, \varphi_{3}  \right\rbrace = 1$. Thus, we infer $\varphi_{2}$ are $ \varphi_{3}$ are second-class in nature. By constructing the non-singular anti-symmetric $2 \times 2$ matrix, $C_{ij} = \left\lbrace \varphi_{i}, \varphi_{j} \right\rbrace$, with $i,j = 2, 3$, we obtain the following Dirac brackets 
\begin{equation} \label{DB}
	\left\lbrace \theta, P_{\theta} \right\rbrace_{D} \;=\; 1 \;=\; \left\lbrace \phi, P_{\phi} \right\rbrace_{D},
\end{equation}
the remaining Dirac brackets are turn out to be zero.

\section{Modified Faddeev-Jackiw formalism}
In order to implement the Faddeev-Jackiw formalism \cite{FJ1}, we express the Lagrangian 
\eqref{L_original} in the first-order form by introducing auxiliary variables. Here, by choosing the canonical momenta as auxiliary variables, the first-order Lagrangian $L_{f}^{(0)}$ can be written in the following fashion 
\begin{equation} \label{L_f_0}
	L_{f}^{(0)} \;=\; \dot{r}P_{r} + \dot{\theta}P_{\theta} + \dot{\phi}P_{\phi} - V^{(0)},
\end{equation}
where the symplectic potential $V^{(0)}$ is given as 
\begin{equation} \label{V0}
	V^{(0)} \;=\; \frac{P_{r}^{2}}{2} + \frac{P_{\theta}^{2}}{2r^{2}} + \frac{P_{\phi}^{2}}{2(R+r \sin\theta)^{2}} - \lambda(r-a).
\end{equation}
We choose the set of zeroth-iterated symplectic variables $\zeta^{(0)}$ corresponding to the first-order Lagrangian \eqref{L_f_0}
as $\zeta^{(0)} = \lbrace r, P_{r}, \theta, P_{\theta}, \phi, P_{\phi}, \lambda \rbrace$. In Faddeev-Jackiw formalism, the Euler-Lagrange equations of motion are expressed in terms of symplectic matrix $f_{ij}^{(0)}$ in the following manner 
\begin{eqnarray} \label{f_ij_eqm}
	f_{ij}^{(0)}\dot{\zeta}^{(0)j} \;=\; \frac{\partial V^{(0)}(\zeta)}{\partial \zeta^{(0)i}},
\end{eqnarray}
where the symplectic matrix is: $f_{ij}^{(0)} = \frac{\partial a_{j}(\zeta)}{\partial \zeta^{i}}- \frac{\partial a_{i}(\zeta)}{\partial \zeta^{j}},$
with canonical one-forms $a_{i} = \frac{\partial L}{\partial \dot{\zeta_{i}}}$. The components of canonical one-form corresponding to the zeroth-iterated symplectic variables are identified as
\begin{equation}
	a_{r}^{(0)} \;=\; P_{r}, \quad a_{\theta}^{(0)} \;=\; P_{\theta}, \quad a_{\phi}^{(0)} \;=\; P_{\phi}, \quad
	a_{P_{r}}^{(0)} \;=\; a_{P_{\theta}}^{(0)} \;=\; a_{P_{\phi}}^{(0)} \;=\; a_{\lambda}^{(0)} \;=\; 0 , 
\end{equation}
with this symplectic matrix $f_{ij}^{(0)}$ takes following form
\begin{equation}
	f_{ij}^{(0)} \;=\;
	\begin{pmatrix}
		0 & -1 & 0 & 0 & 0 & 0 & 0 \\
		1 & 0 & 0 & 0 & 0 & 0 & 0 \\
		0 & 0 & 0 & -1 & 0 & 0 & 0 \\
		0 & 0 & 1 & 0 & 0 & 0 & 0 \\
		0 & 0 & 0 & 0 & 0 & -1 & 0 \\
		0 & 0 & 0 & 0 & 1 & 0 & 0 \\
		0 & 0 & 0 & 0 & 0 & 0 & 0
	\end{pmatrix}.
 \end{equation}
Here we observe $f_{ij}^{(0)}$ is a singular matrix, which in turn indicates the presence of constraints. 
The zero-mode of this singular matrix $f_{ij}^{(0)}$ is $(\nu^{(0)})^{T} = (0 \; 0 \; 0 \; 0 \; 0 \; 0 \; \nu_\lambda) $, where $\nu_\lambda$ is an arbitrary constant. Contracting this zero-mode with the Euler-Lagrange equations produces the constraint $\Omega^{(0)}$ in the system as
\begin{eqnarray} \label{C_0}
	\Omega^{(0)} \;\equiv\; (\nu^{(0)})^{T}\frac{\partial V^{(0)}(\zeta)}{\partial \zeta^{(0)}} \;=\; 0 \quad \Longrightarrow  \quad
	\Omega^{(0)} \; \equiv \; \nu_{\lambda}(r-a) \;=\; 0.
\end{eqnarray} 
Now, we implement modified Faddeev-Jackiw formalism to derive new constraints in the theory \cite{modFJ2}. In the modified Faddeev-Jackiw formalism, the consistency condition of $\Omega^{(0)}$, analogous to Dirac-Bergmann methodology, is used to obtain the new constraint in the system as follows:
\begin{eqnarray}
	\dot{\Omega}^{(0)} \;=\; \frac{\partial{\Omega^{(0)}}}{\partial \zeta^{i}}\dot{\zeta^{i}} \;= \;0 .\label{mfj_c}
\end{eqnarray}
Combining this consistency condition \eqref{mfj_c}  of constraint $\Omega^{(0)}$ and symplectic equations of motion \eqref{f_ij_eqm}, we obtain following relationship (cf. e.g. \cite{modFJ3, modFJ4})
\begin{eqnarray}
	\bar{f}_{kj}^{(0)} \dot{\zeta^{j}} \;=\; Z_{k}^{(0)}(\zeta), \label{mfj_eqm}
\end{eqnarray}
where
\begin{eqnarray}
	\bar{f}_{kj}^{(0)} \;=\;
	\begin{pmatrix}
		f_{ij}^{(0)}\\
		\frac{\partial{\Omega^{(0)}}}{\partial \zeta^{j}}
	\end{pmatrix}
	, \quad Z_{k}^{(0)}(\zeta) \;=\;
	\begin{pmatrix}
		\frac{\partial V^{(0)}(\zeta)}{\partial \zeta^{i}}\\
		0
	\end{pmatrix}.
\end{eqnarray}
The symplectic matrix $\bar{f}_{kj}^{(0)}$ yields
\begin{eqnarray}
	\bar{f}_{kj}^{(0)} \;=\;
	\begin{pmatrix}
		0 & -1 &0 &0 &0 &0 &0\\
		1 &0 &0 &0 &0 &0 &0\\
		0 &0 &0 &-1 &0 &0 &0\\
		0 &0 &1 &0 &0 &0 &0\\
		0 &0 &0 &0 &0 &-1 &0\\
		0 &0 &0 &0 &1 &0 &0\\
		0 &0 &0 &0 &0 &0 &0 \\
		1 &0 &0 &0 &0 &0 &0
	\end{pmatrix},
\end{eqnarray}
which is a non-square matrix and possess a zero-mode of order $1 \times 8$. The zero-mode of $\bar{f}_{kj}^{(0)}$ is given as, 
$(\bar{\nu}^{(0)})^{T}=(0\;1 \;0 \;0 \;0 \;0 \;\nu_{\lambda}^{1}\;-1)$, where $\nu_{\lambda}^{1}$ is an arbitrary constant. Multiplying $(\bar{\nu}^{(0)})^{T}$ to \eqref{mfj_eqm} 
leads to a new constraint in the system with the condition $\Omega^{(0)}=0$. So, we obtain the following relation
\begin{eqnarray} \label{con_omega_0}
	(\bar{\nu}^{(0)})^{T} Z_{k}^{(0)} (\zeta)\vert_{\Omega^{(0)}=0} \;=\; 0  \quad \Longrightarrow \quad  
	 P_{r} - \nu_{\lambda}^{1} (r-a)\vert_{\Omega^{(0)}=0} \;=\; 0,
\end{eqnarray}
which does not vanish identically. This indicates that there exist more constraints in the system. The constraint $\Omega^{(0)}$ is now introduced into the canonical sector of first-order Lagrangian \eqref{L_f_0} with the help of Lagrange multiplier, say $\beta$. Thus, we have 
\begin{equation} \label{L_f_1}
	L_{f}^{(1)} \;=\; \dot{r}P_{r} + \dot{\theta}P_\theta + \dot{\phi}P_\phi +\dot{\beta}(r-a) - V^{(1)},
\end{equation}
with the first-iterated symplectic potential $V^{(1)}$ given by
\begin{equation} \label{V1}
	V^{(1)} \;=\; V^{(0)}\vert_{\Omega^{(0)}=0} \;=\; \frac{P_{r}^{2}}{2} + \frac{P_{\theta}^{2}}{2r^{2}} + \frac{P_{\phi}^{2}}{2(R+r \sin\theta)^{2}}.
\end{equation}
Now, we choose the set of first-iterated symplectic variables as $\zeta^{(1)}= \left\{ r, P_{r}, \theta, P_\theta, \phi, P_\phi, \beta \right\}$. The corresponding symplectic one-forms are identified as
\begin{equation}
	a^{(1)}_{r}\;=\;P_{r},\quad a^{(1)}_{\theta}\;=\;P_{\theta}, \quad a^{(1)}_{\phi}\;=\;P_{\phi}, \quad
	a^{(1)}_{\beta} \;=\; r-a, \quad a^{(1)}_{P_{r}}\;=\; a^{(1)}_{P_{\theta}} \;=\; a^{(1)}_{P_{\phi}} \;=\; 0.
\end{equation}
The first-iterated symplectic matrix $f_{ij}^{(1)}$  takes the form 
\begin{eqnarray}
	f_{ij}^{(1)} \;=\;
	\begin{pmatrix}
		0 & -1 &0 &0 &0 &0 &1\\
		1 &0 &0 &0 &0 &0 &0\\
		0 &0 &0 &-1 &0 &0 &0\\
		0 &0 &1 &0 &0 &0 &0\\
		0 &0 &0 &0 &0 &-1 &0\\
		0 &0 &0 &0 &1 &0 &0\\
		-1 &0 &0 &0 &0 &0 &0
	\end{pmatrix},
\end{eqnarray}
which is a singular matrix. The zero mode 
corresponding to $f_{ij}^{(1)}$ is calculated as $(\nu^{(1)})^{T} = (0 \;1 \;0 \;0 \;0 \;0 \;1)$, which is of order $1 \times 7$. Now, multiplying this zero-mode with the Euler-Lagrange equations, 
the constraint in the theory is deduced as 
\begin{equation} \label{C_1}
	\Omega^{(1)} \; \equiv \; (\nu^{(1)})^{T}\frac{\partial V^{(1)}(\zeta)}{\partial \zeta^{(1)}} \;=\; 0  \quad \Longrightarrow  \quad
	\Omega^{(1)} \; \equiv \; P_{r} \;=\; 0.
\end{equation}
We again make use of modified Faddeev-Jackiw formalism to seek the existence of further constraints. Here we use the consistency condition of $\Omega^{(1)}$ and combine it with the equations of motion produced by the first-iterated Lagrangian $L_{f}^{(1)}$ and we obtain the following expression
\begin{equation} \label{euler_2}
	\bar{f}_{kj}^{(1)} \dot{\zeta^{j}} \;=\; Z_{k}^{(1)}(\zeta),
\end{equation}
where $\bar{f}_{kj}^{(1)}$ and $Z_{k}^{(1)}$ take the following form, respectively 
\begin{eqnarray}
	\bar{f}_{kj}^{(1)} \;=\;
	\begin{pmatrix}
		f_{ij}^{(1)}\\
		\frac{\partial{\Omega^{(1)}}}{\partial \zeta^{j}}
	\end{pmatrix}
	, \quad Z_{k}^{(1)}(\zeta) \;=\;
	\begin{pmatrix}
		\frac{\partial V^{(1)}(\zeta)}{\partial \zeta^{i}}\\
		0
	\end{pmatrix}.
\end{eqnarray}
The symplectic matrix $\bar{f}_{kj}^{(1)}$ is given by
\begin{eqnarray}
	\bar{f}_{kj}^{(1)} \;=\;
	\begin{pmatrix}
		0 & -1 &0 &0 &0 &0 &1\\
		1 &0 &0 &0 &0 &0 &0\\
		0 &0 &0 &-1 &0 &0 &0\\
		0 &0 &1 &0 &0 &0 &0\\
		0 &0 &0 &0 &0 &-1 &0\\
		0 &0 &0 &0 &1 &0 &0\\
		-1 &0 &0 &0 &0 &0 &0 \\
		0 &1 &0 &0 &0 &0 &0
	\end{pmatrix}.
\end{eqnarray}
However the symplectic matrix $\bar{f}_{kj}^{(1)}$ is a non-square one, it still has a zero-mode of order $1 \times 8$, $(\bar{\nu}^{(1)})^{T}=(0 \;1 \;0 \;0 \;0\;0\;1\;0)$. 
The contraction of this zero-mode with $Z_{k}^{(1)}(\zeta)$, by imposing the condition $\Omega^{(1)} = 0$, yields
\begin{equation} \label{mod_1}
	(\bar{\nu}^{(1)})^{T}Z_{k}^{(1)}(\zeta)\vert _{\Omega^{(1)} = 0} \;=\; 0 \quad \Longrightarrow \quad P_{r}\vert _{\Omega^{(1)} = 0}\;=\;0.
\end{equation}
The equation \eqref{mod_1} gives identity, indicating there exist no further constraints in the theory.
Now we construct the second-iterated Lagrangian by incorporating the constraint $\Omega^{(1)}$ into first-iterated Lagrangian \eqref{L_f_1} with the help of Lagrange multiplier, say $\alpha$, as
\begin{equation} \label{L_f_2}
	L_{f}^{(2)} \;=\; \dot{r}P_{r} + \dot{\theta}P_\theta + \dot{\phi}P_\phi +\dot{\beta}(r-a) + \dot{\alpha}P_{r} - V^{(2)},
\end{equation} \label{V2}
where the symplectic potential $V^{(2)}$ is given by
\begin{equation}
	V^{(2)} \;=\; V^{(1)}\vert_{\Omega^{(1)}=0} \;=\; \frac{P_{\theta}^{2}}{2r^{2}} + \frac{P_{\phi}^{2}}{2(R+r \sin\theta)^{2}}.
\end{equation}
Here we choose the  second-iterated symplectic variables as $\zeta^{(2)}= \left\{r, P_{r}, \theta, P_\theta, \phi, P_\phi, \beta, \alpha \right\}.$ The corresponding canonical one-forms are given as below 
\begin{eqnarray}
	&a^{(2)}_{r} \;=\; P_{r}, \quad a^{(2)}_{\theta} \;=\; P_{\theta}, \quad a^{(2)}_{\phi} \;=\; P_{\phi}, \quad a^{(2)}_{\alpha} \;=\; P_{r},& \\ \nonumber
	& a^{(2)}_{\beta} \;=\; r-a, \quad a^{(2)}_{P_{r}} \;=\; a^{(2)}_{P_{\theta}} \;=\; a^{(2)}_{P_{\phi}} \;=\; 0.&
\end{eqnarray}
The second-iterated symplectic matrix $f_{ij}^{(2)}$ takes the form 
\begin{eqnarray}
	f_{ij}^{(2)} \;=\;
	\begin{pmatrix}
		0 & -1 &0 &0 &0 &0 &1 &0\\
		1 &0 &0 &0 &0 &0 &0 &1\\
		0 &0 &0 &-1 &0 &0 &0 &0\\
		0 &0 &1 &0 &0 &0 &0 &0\\
		0 &0 &0 &0 &0 &-1 &0 &0\\
		0 &0 &0 &0 &1 &0 &0 &0\\
		-1 &0 &0 &0 &0 &0 &0 &0 \\
		0 &-1 &0 &0 &0 &0 &0 &0
	\end{pmatrix}.
\end{eqnarray}
We recognize the second-iterated symplectic matrix $f_{ij}^{(2)}$ is a non-singular one, indicating that a free particle system on toric geometry is not a gauge invariant theory. It is worthwhile to mention that in the presence of gauge symmetry the symplectic matrix $f_{ij}^{(2)}$ would have been singular and we would have to implement a suitable gauge condition to obtain the non-singular symplectic matrix \cite{FJ1}.  Since our second-iterated matrix  $f_{ij}^{(2)}$ is non-singular, we can determine all the basic brackets in the theory from its inverse. The inverse of $f_{ij}^{(2)}$ is given by 
\begin{eqnarray} \label{f_ij_2_inverse}
	(f_{ij}^{(2)})^{-1} \;=\;
	\begin{pmatrix}
		0 &0 &0 &0 &0 &0 &-1 &0\\
		0 &0 &0 &0 &0 &0 &0 &-1\\
		0 &0 &0 &1 &0 &0 &0 &0\\
		0 &0 &-1 &0 &0 &0 &0 &0\\
		0 &0 &0 &0 &0 &1 &0 &0\\
		0 &0 &0 &0 &-1 &0 &0 &0\\
		1 &0 &0 &0 &0 &0 &0 &-1 \\
		0 &1 &0 &0 &0 &0 &1 &0
	\end{pmatrix}.
\end{eqnarray}
The basic brackets in the theory can be procured, by  identifying $\lbrace \zeta_{i}^{(2)}, \zeta_{j}^{(2)} \rbrace = (f_{ij}^{(2)})^{-1}$, as
\begin{equation} \label{Basic_brac_FJ}
	 \lbrace \theta, P_{\theta} \rbrace \;=\; \lbrace \phi, P_{\phi} \rbrace \;=\; 1, \quad
	 \lbrace r, \beta \rbrace \;=\; \lbrace P_{r}, \alpha \rbrace \;=\; \lbrace \beta, \alpha \rbrace \;=\; -1. 
\end{equation}
The rest of the brackets are zero. We infer the brackets, among the basic variables, calculated by the modified Faddeev-Jackiw formalism (cf. \eqref{Basic_brac_FJ}) coincide with the corresponding  Dirac brackets (cf. \eqref{DB}).

Before ending this section we would like to mention that the Lagrange multipliers appearing as auxiliary coordinates to include the constraints into the canonical sector of first-order Lagrangian in the Faddeev-Jackiw formalism are absent in the Dirac formalism. So, we provide a {\it{new}} interpretation for these Lagrange multipliers ($\alpha$, $\beta$)  in terms of physical coordinates of the system by making use of  symplectic equations of motion of second-iterated Lagrangian \eqref{L_f_2} as
\begin{equation}
	\dot{\alpha} \;=\; - \dot{r}, \quad
	\dot{\beta} \;=\; \dot{P_{r}} - \frac{P_{\theta}^{2}}{r^{3}} - \frac{P_{\phi}^{2} \sin\theta}{(R+r \sin\theta)^{3}}.
\end{equation}
Thus, the Lagrange multipliers are expressed in terms of generalized coordinates and momenta in the theory.

\section{Symplectic gauge invariant formalism} 
We have inferred in the previous section that the free particle system on a toric geometry is not a gauge invariant theory. We implement the symplectic gauge invariant formalism \cite{4, 5} to reformulate the system as a gauge theory. In this procedure  the key idea is to enlarge the original phase space with the incorporation of Wess-Zumino variable. To accomplish this, we introduce a new function $G$ depending upon the original phase space variables and the 
Wess-Zumino variable, say $\rho$, which expands as
\begin{equation}
	G(r, P_{r}, \theta, P_{\theta}, \phi, P_{\phi}, \rho) \;=\; \sum_{n=0}^{\infty} \mathcal{G}^{(n)} (r, P_{r}, \theta, P_{\theta}, \phi, P_{\phi}, \rho),
\end{equation}
where the quantity $\mathcal{G}^{(n)}$ represents a $n^{th}$ order term in the Wess-Zumino variable $\rho$ which satisfies the following boundary condition
\begin{equation}
	G(r, P_{r}, \theta, P_{\theta}, \phi, P_{\phi}, \rho = 0) \;=\; \mathcal{G}^{(0)} \;=\; 0.
\end{equation}
Introducing this new term $G$ into the first-iterated Lagrangian \eqref{L_f_1}, which now depends on original phase space variables as well as Wess-Zumino variable, as 
\begin{equation} \label{L_f_g_1}
	\tilde{L}_{f}^{(1)} \;=\; \dot{r}P_{r} + \dot{\theta}P_{\theta} + \dot{\phi}P_{\phi} +\dot{\beta}(r-a) - \tilde{V}^{(1)},
\end{equation} 
where the form of symplectic potential $\tilde{V}^{(1)}$ becomes
\begin{equation} \label{V1_g}
	\tilde{V}^{(1)} \;=\; \frac{P_{r}^{2}}{2} + \frac{P_{\theta}^{2}}{2r^{2}} + \frac{P_{\phi}^{2}}{2(R+r \sin\theta)^{2}} - G.
\end{equation}
Correspondingly, the set of symplectic variables is enlarged with the inclusion of Wess-Zumino variable as $\tilde{\zeta}^{(1)}= \left\{r, P_{r}, \theta, P_{\theta}, \phi, P_{\phi}, \beta, \rho \right\}.$
Now from the new first-iterated Lagrangian (cf. \eqref{L_f_g_1}), we identify the symplectic one-forms as
\begin{eqnarray}
	&\tilde{a}^{(1)}_{r} \;=\; P_{r},\quad \tilde{a}^{(1)}_{\theta} \;=\; P_{\theta}, \quad \tilde{a}^{(1)}_{\phi} \;=\; P_{\phi}, &\\ \nonumber
	& \tilde{a}^{(1)}_{\beta} \;=\; r-a, \quad \tilde{a}^{(1)}_{P_{r}} \;=\; \tilde{a}^{(1)}_{P_{\theta}} \;=\; \tilde{a}^{(1)}_{P_{\phi}} \;=\; \tilde{a}^{(1)}_{\rho} \;=\; 0.&
\end{eqnarray}
The corresponding symplectic matrix $\tilde{f}_{ij}^{(1)}$ takes the form 
\begin{eqnarray} \label{f_singular}
	\tilde{f}_{ij}^{(1)} \;=\;
	\begin{pmatrix}
		0 & -1 &0 &0 &0 &0 &1 &0\\
		1 &0 &0 &0 &0 &0 &0 &0\\
		0 &0 &0 &-1 &0 &0 &0 &0\\
		0 &0 &1 &0 &0 &0 &0 &0\\
		0 &0 &0 &0 &0 &-1 &0 &0\\
		0 &0 &0 &0 &1 &0 &0 &0\\
		-1 &0 &0 &0 &0 &0 &0 &0 \\
		0 &0 &0 &0 &0 &0 &0 &0
	\end{pmatrix}.
\end{eqnarray}
Here, we infer the symplectic matrix $\tilde{f}_{ij}^{(1)}$ is singular and has following zero-mode of order $1 \times 8$, as
\begin{equation} \label{nu_gauge}
	(\tilde{\nu}^{(1)})^{T} \;=\; 
	\begin{pmatrix}
		0 &1 &0 &0 &0 &0 &1 &1
	\end{pmatrix}.
\end{equation}
In the symplectic gauge invariant formalism, we impose the condition that the corresponding zero-mode of singular symplectic matrix does not generate any further constraints in the theory. Given this condition, we calculate the first-order correction term, in $\rho$, of the newly introduced function $G$ by the following expression
\begin{equation} \label{g_1}
	(\nu^{(1)})^{T}\frac{\partial V^{(1)}}{\partial \zeta^{(1)}} - \frac{\partial \mathcal{G}^{(1)}}{\partial \rho} \;=\; 0.
\end{equation}
Integrating the above equation \eqref{g_1} leads to the first-order correction term of $G$ as
\begin{equation} \label{g_rho_1}
	\mathcal{G}^{(1)} \;=\; P_{r} \rho.
\end{equation}
Incorporating the first-order correction $\mathcal{G}^{(1)}$ into the first-iterated Lagrangian (cf. \eqref{L_f_g_1}), we obtain
\begin{equation} \label{L_f_g_2}
	\tilde{L'}_{f}^{(1)} \;=\; \dot{r}P_{r} + \dot{\theta}P_{\theta} + \dot{\phi}P_{\phi} +\dot{\beta}(r-a) - \tilde{V'}^{(1)},
\end{equation}
where the explicit form of symplectic potential $\tilde{V'}^{(1)}$ is 
\begin{equation} \label{V_g1}
	\tilde{V'}^{(1)} \;=\; \frac{P_{r}^{2}}{2} + \frac{P_{\theta}^{2}}{2r^{2}} + \frac{P_{\phi}^{2}}{2(R+r \sin\theta)^{2}} - P_{r} \rho.
\end{equation}
We observe that the action of the zero-mode  $(\tilde{\nu}^{(1)})^{T}$ (cf. \eqref{nu_gauge}) on the gradient of new symplectic potential $\tilde{V'}^{(1)}$, i.e. $(\tilde{\nu}^{(1)})^{T} \frac{\partial\tilde{V'}^{(1)}}{\partial \tilde{\zeta}^{(1)}} = 0$, does not generate identity. The presence of non-zero term implies that 
we need to calculate higher order correction terms. Thus, by employing the same condition that $(\tilde{\nu}^{(1)})^{T}$ does not produce any new constraints, we calculate the second-order correction in $\rho$ (i.e. $\mathcal{G}^{(2)}$) from the following expression
\begin{equation}
	(\tilde{\nu}^{(1)})^{T} \frac{\partial\tilde{V'}^{(1)}}{\partial \tilde{\zeta}^{(1)}} - \frac{\partial \mathcal{G}^{(2)}}{\partial \rho} \;=\; 0.
\end{equation}
We obtain the expression for $\mathcal{G}^{(2)}$ as
\begin{equation} \label{g2}
	\mathcal{G}^{(2)} \;=\; - \frac{\rho^{2}}{2}.
\end{equation}
Now, including $\mathcal{G}^{(2)}$ into the extended first-order Lagrangian  $ \tilde{L'}_{f}^{(1)}$ (cf. \eqref{L_f_g_2}), we obtain
\begin{equation} \label{L_f_g_3}
	\tilde{L''}_{f}^{(1)} \;=\; \dot{r}P_{r} + \dot{\theta}P_{\theta} + \dot{\phi}P_{\phi} +\dot{\beta}(r-a) - \tilde{V''}^{(1)},
\end{equation}
where the form of symplectic potential $\tilde{V''}^{(1)}$ is given by
\begin{eqnarray} \label{V_final}
	\tilde{V''}^{(1)} \;=\; \frac{P_{r}^{2}}{2} + \frac{P_{\theta}^{2}}{2r^{2}} + \frac{P_{\phi}^{2}}{2(R+r \sin\theta)^{2}} - P_{r} \rho + \frac{\rho^{2}}{2}.
\end{eqnarray}
Now it is straightforward to check the contraction of zero-mode $(\tilde{\nu}^{(1)})^{T}$ on the gradient 
of symplectic potential $\tilde{V''}^{(1)}$ vanishes identically, thus, this zero-mode does not produce any further constraints. This indicates that the iterative process of calculating higher order correction terms in $G$ with $n\geq 3$ can be stopped. Consequently, the zero mode $(\tilde{\nu}^{(1)})^{T}$ acts as the generator of infinitesimal gauge transformations \cite{4}. 

Thus, with the new symplectic potential \eqref{V_final} obtained via symplectic gauge invariant formalism, we express the zeroth-iterated first-order Lagrangian as
\begin{equation} \label{L_f_g_0}
	\tilde{L}_{f}^{(0)} \;=\; \dot{r}P_{r} + \dot{\theta}P_{\theta} + \dot{\phi}P_{\phi} - \tilde{V}^{(0)},
\end{equation}
where the explicit form of $\tilde{V}^{(0)}$ is given by
\begin{equation} \label{V_0_F}
	\tilde{V}^{(0)} \;=\; \frac{P_{r}^{2}}{2} + \frac{P_{\theta}^{2}}{2r^{2}} + \frac{P_{\phi}^{2}}{2(R+r \sin\theta)^{2}} 
	-\lambda(r-a) - P_{r} \rho + \frac{\rho^{2}}{2} \quad \; \equiv \; \quad \tilde{H}.
\end{equation}
Here we have identified the newly constructed symplectic potential (cf. \eqref{V_0_F}) as the Hamiltonian of the system \cite{4}.
As we have mentioned earlier, the zero mode $(\tilde{\nu}^{(1)})^{T}$ acts as the generator of gauge symmetry, we can compute the gauge transformations ($\delta$) from the following expression
\begin{equation}
	\delta \tilde{\zeta}^{(1)} \;=\; \epsilon (\tilde{\nu}^{(1)})^{T},
\end{equation}
where $\tilde{\zeta}^{(1)}$ is the set of  symplectic variables in the enlarged phase space and $\epsilon$ represents the infinitesimal time dependent gauge parameter. The explicit form of non-zero infinitesimal gauge transformations in the theory are
\begin{equation} \label{gau_trans}
	\delta P_{r} \;=\; \epsilon, \qquad \delta \lambda \;=\; \dot{\epsilon}, \qquad \delta \rho \;=\; \epsilon.
\end{equation}
 Under these gauge transformations \eqref{gau_trans},  the first-order Lagrangian \eqref{L_f_g_0}  transforms as
\begin{equation}
	\delta \tilde{L}_{f}^{(0)}  \;=\; \frac{d}{dt} \Big[  \epsilon (r-a) \Big] .
\end{equation}
Now, we wish to rewrite the gauge invariant Lagrangian \eqref{L_f_g_0} and corresponding Hamiltonian \eqref{V_0_F} in terms of the original phase space variables. By employing Dirac formalism, we identify conjugate momenta corresponding to the variables $\lambda$ and $\rho$ as the primary constraints and they generate two sets of constraint chain in the theory as given below, respectively
\begin{equation}
	\psi_{1} \;=\; P_{\lambda}, \quad
	\psi_{2} \;=\; (r-a),
\end{equation}
and
\begin{equation}
	\bar{\psi}_{1} \;=\; P_{\rho}, \quad
	\bar{\psi}_{2} \;=\; P_{r} - \rho,
\end{equation}
where $P_{\rho}$ is the canonical conjugate momentum to Wess-Zumino variable $\rho$. It is worthwhile to mention few points in order, in the first constraint chain generated by $\lambda$, the consistency condition of $\psi_{2}$ leads to the tertiary constraint ($\psi_{3} = P_{r} - \rho$). Since this constraint is not unique, we have the option of putting this constraint in either one of the chain and close the other chain at the same time \cite{mit1}. So, we have put this constraint in second chain and closed the first one. Further we observe there are non-vanishing Poisson brackets between the constraints, as $\left\lbrace \psi_{2}, \bar{\psi}_{2} \right\rbrace = 1 = \left\lbrace \bar{\psi}_{1}, \bar{\psi}_{2} \right\rbrace$. This indicates that there is a mixture of both first-class and second-class constraints. 
We split the constraints into first-class and second-class through constraint combination \cite{4}. The set of first-class constraints takes the following form
\begin{equation} \label{firstclass_f}
	\chi_{1} \;=\; P_{\lambda}, \quad
	\chi_{2} \;=\; r-a-P_{\rho},
\end{equation}
and the set of second-class constraints is
\begin{eqnarray}
	\bar{\psi}_{1} \;=\; P_{\rho}, \quad
	\bar{\psi}_{2} \;=\; P_{r} - \rho.
\end{eqnarray}
Now assuming the second-class constraints in a strong way, we rewrite the Hamiltonian  \eqref{V_0_F} in terms of the original phase space variables as
\begin{eqnarray} \label{H'_fin}
	\tilde{H'} \;=\; \frac{P_{\theta}^{2}}{2r^{2}} + \frac{P_{\phi}^{2}}{2(R+r \sin\theta)^{2}} -\lambda(r-a).
\end{eqnarray}

We wish to disclose the gauge symmetry existing in the gauge theory developed by symplectic gauge invariant formalism. Thus, expressing the first-order Lagrangian corresponding to the final Hamiltonian \eqref{H'_fin} as 
\begin{equation} \label{L'_final}
	L_{f} \;=\; \dot{r}P_{r} + \dot{\theta}P_{\theta} + \dot{\phi}P_{\phi} 
	- \frac{P_{\theta}^{2}}{2r^{2}} - \frac{P_{\phi}^{2}}{2(R+r \sin\theta)^{2}} + \lambda(r-a).
\end{equation}
The set of constraints in the theory is obtained as
\begin{equation} \label{first_class}
	\bar{\chi}_{1} \;=\; P_{\lambda}, \quad
		 \bar{\chi}_{2} \;=\; (r-a). 
\end{equation}
The constraints $\bar{\chi}_{1}$ and $\bar{\chi}_{2}$ are first-class in nature and act as the generators of gauge symmetry. The corresponding gauge symmetry transformations can be given as
\begin{equation} \label{gauge_trans_firstclass}
	\delta \lambda \;=\; \dot{\kappa}, \quad \delta P_{r} \;=\; \kappa, \quad \delta[r,\; \theta, \; P_{\theta}, \; \phi,  \; P_{\phi}] \;=\; 0,
\end{equation}
where $\kappa$ is the infinitesimal time dependent gauge parameter. Under these gauge transformations in \eqref{gauge_trans_firstclass}, the first-order Lagrangian \eqref{L'_final} transforms as
\begin{equation}
	\delta L_{f} \;=\; \frac{d}{dt} \Big[ \kappa(r-a) \Big] .
\end{equation}
 
At this juncture, we would like to mention some salient features of this reformulation. First and foremost, although we have started with the inclusion of Wess-Zumino term in the gauge non-invariant Lagrangian (cf. \eqref{L_f_1}), our final gauge invariant Lagrangian does not contain any additional variables other than the original phase space variables (cf. \eqref{L'_final}). Moreover, the first-class constraints in the gauge invariant theory retain the same form as of the original system. Second, the reformulated gauge invariant theory is equivalent to the original gauge non-invariant theory 
in the sense that the physical degrees of freedom and dynamics remain same in both the cases.  The physical degrees of freedom can be checked  using constraints. To be specific, there is only one physical degree of freedom present in original gauge non-invariant theory as it contains one first-class constraint and two second-class constraints. On the other hand the reformulated gauge theory, consisting of two first-class constraints, also has only one degree of freedom.  In addition, it can be verified in a straightforward manner that the  dynamical equations arising from the gauge non-invariant theory are same as the equations obtained from the newly constructed gauge theory.

\section{Off-shell nilpotent and absolutely anti-commuting (anti-)BRST symmetries}
We wish to investigate the reformulated gauge theory of a free particle system on  toric geometry in the BRST framework. To accomplish this we replace the time dependent gauge parameter by anti-commuting (anti-)ghost variables ($\bar{c})c$, satisfying $c^{2} = 0 = \bar{c}^{2}, c\bar{c}+\bar{c}c=0$. The BRST invariant first-order Lagrangian ($L_{f}^{(b)}$) can be obtained by adding a BRST invariant function $ s_{b} \left(- i\bar{c}\left(-\dot{\lambda} - P_{r} + \frac{1}{2}b \right)  \right)$ to the first-order Lagrangian \eqref{L'_final}, which respects the following BRST transformations $s_{b}$ 
\begin{eqnarray} \label{brst}
&s_{b}\lambda \;=\; \dot{c}, \quad 	s_{b} P_{r} \;=\; c, \quad 	s_{b}\bar{c} \;=\; ib, \quad	s_{b} c \;=\; 0  ,& \\ \nonumber
	&s_{b} b \;=\; s_{b} r \;=\;  s_{b} \theta \;=\;  s_{b} P_{\theta} \;=\;  s_{b} \phi \;=\; s_{b} P_{\phi}  \;=\; 0. &
\end{eqnarray}
Thus, the BRST invariant first-order Lagrangian $L_{f}^{(b)}$ is given as
\begin{eqnarray} \label{L_brst_1}
	L_{f}^{(b)} &=& \dot{r}P_{r} + \dot{\theta}P_{\theta} + \dot{\phi}P_{\phi} - \frac{P_{\theta}^{2}}{2r^{2}}  - \frac{P_{\phi}^{2}}{2(R+r \sin\theta)^{2}} + \lambda(r-a) \\ \nonumber
	&-&  b(\dot{\lambda} + P_{r}) + \frac{1}{2}b^{2} + i\dot{\bar{c}}\dot{c} -i\bar{c}c.
\end{eqnarray}
In the above expression, $b$ is Nakanishi-Lautrup auxiliary variable which linearizes the gauge fixing condition $-\frac{1}{2}(\dot{\lambda} + P_{r})^{2}$. The first-order Lagrangian $L_{f}^{(b)}$ is also found to be invariant under another set of symmetries, the anti-BRST transformations $s_{ab}$, given as  
\begin{eqnarray} \label{antibrst}
&	s_{ab}\lambda \;=\; \dot{\bar{c}}, \quad 	s_{ab} P_{r} \;=\; \bar{c}, \quad 	s_{ab}{c} \;=\; -ib, \quad	s_{ab} \bar{c} \;=\; 0 , &\\ \nonumber
&s_{ab} b \;=\;	s_{ab} r \;=\; s_{ab} \theta \;=\;  s_{ab} P_{\theta} \;=\;  s_{ab} \phi \;=\; s_{ab} P_{\phi}  \;=\; 0.&
\end{eqnarray}
These (anti-)BRST transformations are off-shell nilpotent (i.e. $s_{b}^{2} = 0 = s_{ab}^{2} $) and absolutely anti-commuting (i.e. $\left\lbrace s_{b}, s_{ab} \right\rbrace  =  s_{b}s_{ab}+s_{ab}s_{b} = 0$) in nature. The conserved (anti-)BRST charges $Q_{(a)b}$ corresponding to the above mentioned (anti-)BRST symmetries are listed as
\begin{equation}
	Q_{b} \;=\; -c(r-a) + \dot{c}P_{\lambda}, \quad Q_{ab} \;=\; -\bar{c}(r-a) + \dot{\bar{c}}P_{\lambda}.
\end{equation}
Here, we have identified $b$ with $-P_{\lambda}$. The conservation of the (anti-)BRST charges $(\dot{Q}_{(a)b}=0)$ can be verified with the aid of following Euler-Lagrange equations of motion corresponding to $L_{f}^{(b)}$ 
\begin{eqnarray}
	&\dot{P_{r}} \;=\; \displaystyle \frac{P_{\theta}^{2}}{r^{3}} + \frac{P_{\phi}^{2}\sin\theta}{(R+r\sin\theta)^{3}} + \lambda, \quad \dot{P_{\theta}} \;=\;  \frac{P_{\phi}^{2}r\cos\theta}{(R+r\sin\theta)^{3}},  &  \\ \nonumber
&\dot{P_{\phi}}\;=\; 0, \quad  \dot{\theta} \;=\;  \displaystyle \frac{P_{\theta}}{r^{2}}, \quad  \dot{r} \;=\; b, \quad \dot{\phi} \;=\; \displaystyle  \frac{P_{\phi}}{(R+r\sin\theta)^{2}} ,& \\ \nonumber
	&  	b \;=\; \dot{\lambda} + P_{r}, \quad \dot{b}\;=\;-(r-a), \quad \ddot{c}+c\;=\;0, \quad \ddot{\bar{c}}+\bar{c}\;=\;0.&
\end{eqnarray}
These (anti-)BRST charges are nilpotent ($Q_{b}^{2} = 0 = Q_{ab}^{2}$), absolutely anti-commuting ($Q_{b}Q_{ab}+Q_{ab}Q_{b} =0$) in nature and turn out to be the generators of (anti-)BRST transformations (listed in \eqref{brst} and \eqref{antibrst}). Explicitly, it can be  verified from the relation $s_{(a)b}\Phi = i\left[ Q_{(a)b}, \Phi \right] _{\pm}$, with the use of $\left[ r, P_{r}\right] = i = \left[ \lambda, P_{\lambda}\right] $ and $\left\lbrace \dot{\bar{c}}, c \right\rbrace =1, \left\lbrace \dot{c}, \bar{c} \right\rbrace =-1 $, where $\Phi$ is any general variable of the system and $\pm$ on the subscript represent the  (anti-) commutation relations.

Finally, we would like to mention the dynamically stable subspace satisfying the physicality condition, $Q_{(a)b} \left| \varPsi \right\rangle_{phys}   =0 $, leads to $(r-a)\left| \varPsi \right\rangle _{phys} =0$ and $P_{\lambda}\left| \varPsi\right\rangle_{phys}  =0$. These conditions establish that the physical states of the theory are annihilated by the first-class constraints $P_{\lambda}$ and $(r-a)$ which is consistent with the Dirac method of quantization for constrained system.

\section{Conclusions}
In our present investigation, we have deduced the constraint structure and eventually quantized a free particle system residing on a torus satisfying the geometric constraint $(r-a)=0$ within the framework of modified Faddeev-Jackiw formalism. For this purpose, we have included this geometric constraint into the Lagrangian \eqref{L_original} with the aid of Lagrange multiplier unlike \cite{6}. The presence of a non-singular symplectic matrix, after incorporation of all the constraints into the canonical sector of first-order Lagrangian, indicates that the system is gauge non-invariant and thus all the basic brackets have been procured from its inverse \eqref{Basic_brac_FJ}. In addition, we have provided a {\it{new}} interpretation for the Lagrange multipliers in terms of original phase space variables in the theory. 

Furthermore, we have reformulated the system as a gauge theory, which is {\it physically} equivalent to the original gauge non-invariant theory, by employing symplectic gauge invariant formalism. To accomplish this, we have introduced a new function depending on the original phase space variables and Wess-Zumino variable into the first-iterated symplectic potential (cf. \eqref{V1_g}) and imposed the condition that zero-mode of the corresponding singular symplectic matrix does not generate any new constraints. Consequently, we have shown the first-order Lagrangian constructed via symplectic gauge invariant formalism is invariant under the transformations generated by the set of first-class constraints. One of the striking features of our work is the reformulated gauge invariant theory is free from Wess-Zumino variables unlike \cite{6, 1, hodge}.

Finally, we have constructed off-shell nilpotent and absolutely anti-commuting (anti-) BRST symmetries for reformulated gauge invariant theory. We have shown that the conserved (anti-)BRST charges are the generators of (anti-)BRST symmetry transformations. The physicality condition satisfied by these charges shows that the first-class constraints annihilate the physical states of the system and it is on par with the Dirac quantization procedure \cite{int4}.
For future endeavor, it will be an interesting venture to explore a particle moving on torus knot within this framework and the results of the same shall be reported elsewhere soon \cite{torus_knot}. \\

\noindent 
{\Large \bf Acknowledgments:} The support from FRG scheme of National Institute of Technology Calicut is thankfully acknowledged.

\end{document}